\documentstyle{article}

\newlength{\extraspace}
\newlength{\extraspaces}
\newcommand{\be}{\begin{equation}
\addtolength{\abovedisplayskip}{\extraspaces}
\addtolength{\belowdisplayskip}{\extraspaces}
\addtolength{\abovedisplayshortskip}{\extraspace}
\addtolength{\belowdisplayshortskip}{\extraspace}}
\newcommand{\ee}{\end{equation}}

\newcommand{\ba}{\begin{eqnarray}
\addtolength{\abovedisplayskip}{\extraspaces}
\addtolength{\belowdisplayskip}{\extraspaces}
\addtolength{\abovedisplayshortskip}{\extraspace}
\addtolength{\belowdisplayshortskip}{\extraspace}}
\newcommand{\ea}{\end{eqnarray}}

\newcommand{\newsection}[1]{
\vspace{7mm}
\pagebreak[3]
\addtocounter{section}{1}
\setcounter{equation}{0}
\setcounter{subsection}{0}
\large {\bf \thesection. #1}
\nopagebreak
\medskip
\nopagebreak
\hspace{3mm}}

\newcommand{\nonu}{\nonumber \\[.5mm]} 

\begin{document}       

\begin{large}
\centerline{\bf On the link between Shr\"odinger and Vlasov equations}
\end{large}

\centerline{Tigran Aivazian \footnote{Please use email 
tigran@zetnet.co.uk for correspondence}}

\centerline{\bf Abstract}

It is shown that well-known Vlasov equation can be derived by adding "hidden" 
degrees of freedom and subsequent quantization. The Shr\"odinger 
equation obtained in this manner coincides (in x-representation) with the 
kinetic equation for the original dynamical system

\newsection{Two ways to derive the Vlasov equation}

Consider a non-autonomous dynamical system in $R^n$:
\be
{dx^i\over{dt}}=X^i(x^1,...,x^n,t)
\ee
We assume that the vector field X is solenoidal:
\be
divX\equiv{}{\partial{X^i}\over{\partial{x^i}}}=0
\ee
One can derive Vlasov equation (known also as Liouville or collisionless
Boltzmann) from the condition that the integral of the distribution function
f(x,t) is invariant with respect to the phase flow generated by the field X:
\be
\int_{G}f(x,0)d^nx=\int_{g^t(G)}f(x,t)d^nx,
\ee
where $G\subseteq{R^n}$ is a open subset in $R^n$, $g^t(x)$ is a trajectory 
starting at point $x\in{G}$ and $d^nx$ is a standard integration measure in 
Euclidean space. The Vlasov equation has the following form:
\be
{\partial{f}\over \partial{t}}+{\partial \over \partial x^i} \{X^if\}=0
\ee

{\bf Statement}. Kinetic equation can be also obtained by doubling the 
dimension of the base phase space and subsequent quantization of the resulting 
hamiltonian system. Indeed, one can add n variables $p_i$ and a function 
H(x,p,t) so that the dynamical system becomes hamiltonian:
\ba
\dot x^i={\partial H \over \partial p_i}=X^i(x,t),\nonu
\dot p_i=-{\partial H \over \partial x^i}=\
-p_k{\partial X^k(x,t) \over \partial x^i},\nonu
H(x,p,t)=p_iX^i(x,t)
\ea
Since this system is hamiltonian it is possible to construct the corresponding
equations of quantum mechanis for it:
\be
i{\partial{\psi}\over{\partial{t}}}=\hat{H}\psi{}
\ee
Using the condition of divergenless of X we rewrite the last equation as:
\be
{\partial \psi \over \partial t}+X^i{\partial \psi \over \partial x^i}=0,
\ee
As we see the Shr\"odinger equation for wave function $\psi(x,t)$ in 
x-representation coincides with the Vlasov equation for distribution function
f(x,t) which proves our original statement. 

\end{document}